\documentclass[seceq]{ptptex}
\usepackage{amsmath,amssymb}
\usepackage[dvips]{graphicx}
\usepackage{bm}

\DeclareMathOperator{\tr}{tr}

\DeclareMathOperator{\ring}{ring}

\newcommand{\Slash}[1]{{\ooalign{\hfil/\hfil\crcr$#1$}}}
\newcommand{\uni}{\leavevmode\hbox{\small1\normalsize\kern-.33em1}}

\notypesetlogo                       %comment in if to eliminate PTPTeX 
\title{%        %You can use \\ for explicit line-break
An analytic study of the tricritical line in the U($N_f$)$\times$U($N_f$) sigma model%
}

\author{%       %Use \scshape  for the family name
Junichiro \textsc{Yasuda}%
}

\inst{%     %Affiliation, neglected when [addenda] or [errata]
Institute of  Physics, University of Tokyo\ Tokyo 153-8092, Japan}

\abst{%         %this abstract is neglected when [addenda] or [errata]
The tricritical line of the first-order chiral phase transition is investigated in the U($N_f$)$\times$U($N_f$) sigma model by means of the ring improved one-loop finite temperature effective potential. To locate the tricritical line in the space of the coupling constants, we expand the effective potential up to third order in the high temperature expansion, and up to sixth order in the order parameter expansion. In this approximation, the tricritical line can be evaluated to the lowest order of the coupling constants and it follows an analytic relation between the tree-level masses of the scalar bosons. The validity of the high temperature expansion, the order parameter expansion, and the ring improved perturbation is critically examined. This result does not alter if one includes the massless fermions with the small Yukawa couplings. 
}

\begin{document}

\maketitle

\section{Introduction}

% the purpose of the study for the CEP
A tricritical point (or line) is where a first-order phase transition terminates in the phase diagram. It is important to locate the tricritical point in various phenomenological models, since the first-order phase transition of the early universe may produce the baryon asymmetry \cite{Kuzmin:1985mm,Cohen:1990py,Cohen:1990it} or black holes \cite{Kapusta:2007dn} etc. The tricritical point has been investigated in the context of the electroweak phase transition \cite{Kajantie:1996mn,Kajantie:1996qd,Csikor:1998eu,Rummukainen:1998as,Aoki:1999fi} and also in the context of the QCD chiral phase transition at finite temperature and density. \cite{Casalbuoni:2006rs} 

% approach to analyze a CEP until now
There are several methods to analyze a tricritical point. Among these methods, the most reliable one is the numerical simulation. This approach can solve the problem of the infrared singularities from the transverse gauge bosons which are incalculable in the perturbation theory.\cite{Arnold:1992rz} \
%The main approach to analyze a tricritical point have been the numerical simulations, because there is an infrared sensitivity from the gauge bosons which is incalculable in the perturbation theory. 
On the other hand, several authors applied the alternative methods to analyze a tricritical point, for example, the $\varepsilon$-expansion techniques \cite{Gleiser:1992ch,Arnold:1993bq} or the auxiliary mass method \cite{Ogure:1998xu} etc.

% method to analyze a CEP in this paper
%what is the U($N_f$)$\times$U($N_f$) sigma model?
In this paper, we consider another analytic method to analyze a tricritical point by means of the ring improved one-loop finite temperature effective potential. To locate the tricritical point, we firstly expand the effective potential up to third order in the high temperature expansion. Then we expand the effective potential up to sixth order in the order parameter expansion around the tricritical point where an order parameter is small compared to the critical temperature. We apply this method in the U($N_f$)$\times$U($N_f$) sigma model.\footnote{Pisarski and Wilczek have discussed this model as a low energy effective theory of QCD and have shown that if $N_f\geq 3$, the restoration of the chiral symmetry at finite temperature should be first-order.\cite{Pisarski:1983ms} \ Following this argument, several authors have examined the strength of the first-order phase transition in the U($N_f$)$\times$U($N_f$) sigma model from the point of view of the electroweak baryonegesis. \cite{Appelquist:1995en,Khlebnikov:1995qb,Kikukawa:2007zk}\ Although they have introduced the U$_1$(A) breaking terms, we does not include these terms, since in the following analysis we consider the large $N_f$ region where the U$_1$(A) breaking terms are irrelevant for the critical behavior.}\  In our approximation, the tricritical line in the space of the coupling constants of the sigma model can be evaluated to the lowest order of the coupling constants and it follows an analytic relation between the tree-level masses of the scalar bosons. \footnote{We locate the tricritical point related to the diagram T vs. coupling constants not to T vs. baryonic chemical potential.}

% the last work expanding the effective potential
The technique expanding the effective potential by the order parameter up to sixth order has been applied to study the tricritical point of QCD at finite temperature and density.\cite{Barducci:1989wi}\  Their calculation was done by using the two-loop Cornwall, Jackiw and Tomboulis effective potential.\cite{Cornwall:1974vz} \ Although their calculation contains the numerical evaluation of the effective potential, our technique for the sigma model is the purely analytic one and we can derive a relation between the tree-level masses of the scalar bosons on the tricritical line.

% application to the model with Yukawa interaction
%Our analysis is reliable if we introduce the Yukawa interaction. However, 

% contents of this paper
This paper is organized as follows. 
In section~\ref{sec:model}, we describe the U($N_f$) $\times$U($N_f$) sigma model with Yukawa interaction and the effective masses (field dependent mass) of the sigma  model.
In section~\ref{sec:eff}, we describe the ring improved one-loop finite temperature effective potential and the $T$-dependent effective masses of the sigma model.
In section~\ref{sec:cep}, we analyze the tricritical line of the sigma model and examine the validity of our approximations. 
Section~\ref{sec:sd} is devoted to a summary and discussions.

%%%%%%%%%%%%%%%%%%%%%%%%%%%%%%%%%%%%%%%%%%%%%%%
\section{U($N_f$)$\times$U($N_f$) sigma model with Yukawa interaction}
\label{sec:model}

In this paper, we consider the following Lagrangian:
\begin{eqnarray}
\mathcal{L}=\tr |\partial _\mu \Phi|^2+i\bar{\psi}_L^a\Slash{\partial}\psi_L^a+i\bar{\psi}_R^a\Slash{\partial}\psi_R^a-(y\bar{\psi}^a_L\Phi^{ab}\psi^b_R+h.c.)+\mathcal{L}_\Phi,
\label{eq:LlsmKIN}
\end{eqnarray}
where $\mathcal{L}_\Phi$ is the scalar potential term which is written as, 
\begin{align}
\mathcal{L}_\Phi &= 
 - m_{\Phi}^{2} \tr\Phi ^\dagger \Phi 
 - \frac{\lambda _1}{2} (\tr\Phi ^\dagger \Phi)^2 - \frac{\lambda _2}{2}\tr (\Phi ^\dagger \Phi )^2 . 
\label{eq:Llsm}
\end{align}
The field $\Phi(x)$ is an $N_f \times N_f$ complex matrix, and $\psi_{L}^a$ ($\psi_{R}^a$) is the left (right) hand Wyle spinors of $N_f$ flavors $(a=1,2,...,N_f)$ which transform under the chiral symmetry as 
\begin{equation}
\Phi \rightarrow g_L \Phi g_{R}^{-1} ,\quad  
\psi_{L(R)}\rightarrow g_{L(R)}\psi_{L(R)},\quad 
g_L, g_R   \in \text{U($N_f$)}. 
\end{equation}
For stability,  the quartic couplings should satisfy the following conditions at tree-level:
$\lambda_2>0,\lambda_1+\lambda_2/N_f>0$. 

The vacuum expectation value (VEV) of scalar field $\Phi(x)$ is assumed as,
\begin{equation}
\langle \Phi \rangle = \frac{\phi_0}{\sqrt{2N_f}} \uni,
\label{VEV}
\end{equation}
where $\uni$\  
is the $N_f \times N_f$ unit matrix. 
At tree-level, $\phi_0$ is determined by the potential,
\begin{equation}
\label{eq:eff-potential-tree}
V_0(\phi)=\frac{1}{2}m_\Phi^2\phi^2+\frac{1}{8}\left( \lambda_1+\frac{\lambda_2}{N_f} \right)\phi^4 . 
\end{equation}
For $m_{\Phi }^2 < 0$, it is given by 
\begin{equation}
\phi_0=\sqrt{\frac{-2m_\Phi^2}{\lambda_1+\lambda_2/N_f}}. 
\end{equation}
The scalar field $\Phi$ is parameterized around the VEV as follows:
\begin{equation}
\Phi(x) = \frac{ \phi + h + i \eta }{\sqrt{2N_f}}\, \uni
+ \sum_{\alpha=1}^{N_f^2-1} ( \xi^\alpha + i \pi^\alpha ) T^\alpha , 
\end{equation}
where $T^\alpha$ $(\alpha=1,\cdots, N_f^2-1)$ are the generator of SU($N_f$) satisfying the normalization,  
${\rm Tr} (T^\alpha T^\beta)=\delta^{\alpha \beta}/2$. The fields $h, \eta, \xi^\alpha, \pi^\alpha, \psi^a$ acquire masses at the tree-level as summarized in 
Table~\ref{tab:lsmm}, where, 
for notational simplicity, we use the following abbreviations:
\begin{eqnarray}
a_h&=&\frac{3}{2}(\lambda_1+\lambda_2/N_f), \\
a_\xi &=&\frac{1}{2}(\lambda_1+3\lambda_2/N_f),  \\
a_\eta&=&a_\pi=\frac{1}{2}(\lambda_1+\lambda_2/N_f), 
\end{eqnarray}
and
\begin{eqnarray}
b_h &=& a_h - a_\pi = (\lambda_1+\lambda_2/N_f), \\
b_\xi &=& a_\xi - a_\pi=(\lambda_2/N_f). 
\end{eqnarray}
The effective masses and the degrees of freedom of the each fields which we use for the effective potential are also summarized in Table~\ref{tab:lsmm}.
\begin{table}[t]
\caption{The effective masses, the tree-level masses and the degrees of freedom of the fields in 
U($N_f$)$\times$U($N_f$) linear sigma model with Yukawa interaction.}
\begin{center}
  \begin{tabular}{cccc} \hline \hline
    field & $m_i^2(\phi)$ & $m_i^2(\phi_0)$ & $n_i$ \\ \hline
    h & $m_\Phi^2+a_h \phi^2$ & $b_h \phi_0^2$ & 1 \\
    $\xi$ & $m_\Phi^2+a_\xi \phi^2$ & $b_\xi \phi_0^2$ & $N_f^2-1$ \\
    $\eta$ & $m_\Phi^2+a_\eta \phi^2$ & 0 & 1 \\ 
    $\pi$ & $m_\Phi^2+a_\pi \phi^2$ & 0 & $N_f^2-1$ \\ \hline
    $\psi$ & $\frac{1}{2N_f}y^2\phi^2$& $\frac{1}{2N_f}y^2\phi_0^2$ & $-4N_f$ \\ \hline
\end{tabular}
\label{tab:lsmm}
\end{center}
\end{table}
%

%%%%%%%%%%%%%%%%%%%%%%%%%%%%%%%%%%%%%%%%%%%%%%%
\section{Effective potential}
\label{sec:eff}

In this section, we describe the ring improved one-loop finite temperature effective potential in the U($N_f$)$\times$U($N_f$) sigma model with Yukawa interaction.\cite{Arnold:1992rz,Dolan:1973qd,Carrington:1991hz}\ At tree-level, the potential is given by Eq.(\ref{eq:eff-potential-tree}).
The one-loop contribution at zero temperature, $V_1^{(0)}$, is given by 
\begin{align}
V_1^{(0)}(\phi) =&\sum_{i=s,f}n_i\frac{m_i^4(\phi)}{64\pi^2}
             \left[\ln\frac{m_i^2(\phi)}{\mu^2}-\frac{3}{2}\right] 
             \label{eq:1loop0} , 
\end{align}
in $\overline{\rm MS}$ scheme, where $i$ runs over all of the scalar bosons: $s=\{h, \eta, \xi^\alpha, \pi^\alpha \}$ and the fermions: $f=\{\psi^a\}$. $n_i$ and $m_i(\phi)$  are the number of degrees of freedom and the effective masses depending on $\phi$, respectively. 

The one-loop contribution at finite temperature, $V_1^{(T)}$, is given by 
\begin{align}
V_1^{(T)}(\phi,T) =\frac{T^4}{2\pi^2}\left(\sum_{i=s}n_iJ_{B}[m_i^2(\phi)/T^2]+\sum_{i=f}n_iJ_{F}[m_i^2(\phi)/T^2]\right) , 
\end{align} 
where $J_B$ and $J_F$ are defined by  
\begin{eqnarray}
J_B(a)&=&\int_{0}^{\infty}dx~x^2 \ln\left(1-e^{-\sqrt{x^2+a}}\right), \nonumber \\
J_F(a)&=&\int_{0}^{\infty}dx~x^2 \ln\left(1+e^{-\sqrt{x^2+a}}\right).
\end{eqnarray}
In the high temperature limit where $m(\phi)/T \lesssim 1$,  $J_B$ and $J_F$ can be expanded as follows:
\begin{align}
J_B(m^2/T^2)= 
&-\frac{\pi^4}{45}+\frac{\pi^2}{12}\frac{m^2}{T^2}-\frac{\pi}{6}\left(\frac{m^2}{T^2}\right)^{3/2} \notag \\
&-\frac{1}{32}\frac{m^4}{T^4}\ln\frac{m^2}{a_bT^2}+\mathcal{O}\left(\frac{m^6}{T^6}\right),  \\
J_F(m^2/T^2)= 
&-\frac{7\pi^4}{360}-\frac{\pi^2}{24}\frac{m^2}{T^2} \notag \\
&-\frac{1}{32}\frac{m^4}{T^4}\ln\frac{m^2}{a_fT^2}+\mathcal{O}\left(\frac{m^6}{T^6}\right),  
\end{align}
where $a_b=16\pi^2\exp(3/2-2\gamma_E)$($\ln a_b\approx 5.4076$), $a_f=\pi^2\exp(3/2-2\gamma_E)$($\ln a_f\approx 2.6351$).

One can include the contribution of ring diagrams,  $V_{\ring}(\phi,T)$,  by replacing $m_i^2(\phi)$ in $V_1^{(0)}$ and $V_1^{(T)}$
with the $T$-dependent effective masses $\mathcal{M}_i^2(\phi,T)\equiv m_i^2(\phi)+\Pi_i$,
where $\Pi_i$ is the self-energy of a field $i$ in the IR limit where the Matsubara frequency and the momentum of the external line goes to zero 
and in the leading order of $m_i(\phi)/T$. 

For all of scalar bosons of the sigma model, the self-energies are given by 
\begin{align}
\Pi_s&=\Pi^{(s)}+\Pi^{(f)}, \notag\\
\Pi^{(s)}&=\frac{1}{12}[(N_f^2+1)\lambda_1+2N_f\lambda_2]T^2,\notag\\
\Pi^{(f)}&=\frac{1}{12}y^2T^2, 
\label{eq:self_s}
\end{align}
which are corresponding to the contributions from the scalar bosons themselves and the fermions, respectively. Since fermions do not receive the correction from the ring diagrams, $\Pi_\psi=0$. The $T$-dependent effective masses for scalar bosons and fermions are given by 
\begin{align}
{\cal M}_{s,i}^2(\phi,T)&=m_{\Phi}^2+a_i\phi^2+\left[(N_f^2+1)\lambda_1+2N_f\lambda_2+y^2\right]\frac{T^2}{12} \nonumber \\
&\equiv m_{\Phi}^2+a_i\phi^2+b\frac{T^2}{12},  \nonumber \\
{\cal M}_{f}^2(\phi,T)&=m^2_{\psi}(\phi).
\end{align}

After all, the one-loop ring-improved effective potential is summarized as follows:
\begin{align}
V(\phi) =&V_0(\phi) +V_1^{(0)}(\phi) +V_1^{(T)}(\phi,T)+ V_{\ring}(\phi,T)  \notag \\
             =&V_0(\phi)+\sum_{i=s,f}n_i\frac{\mathcal{M}_i^4(\phi,T)}{64\pi^2}
             \left[\ln\frac{\mathcal{M}_i^2(\phi,T)}{\mu^2}-\frac{3}{2}\right]  \notag \\
          +&\frac{T^4}{2\pi^2}\left(\sum_{i=s}n_iJ_B[\mathcal{M}_i^2(\phi,T)/T^2]+\sum_{i=f}n_iJ_F[\mathcal{M}_i^2(\phi,T)/T^2]\right).
\end{align}

A comment on the validity of the ring-improved perturbation theory is in order. By inspecting the higher order diagrams for the scalar field self-energies, one can see that the non-ring diagrams are suppressed with respect to the ring diagrams at least 
by the following factors in the symmetric phase, \cite{Quiros:1999jp}
\begin{eqnarray}
\beta_{\lambda_1+\lambda_2/N_f}&\equiv& \frac{1}{4\pi}\left(\lambda_1+\frac{\lambda_2}{N_f}\right)\frac{T}{m_{\rm eff}} ,
\label{ring1}\\
\beta_{\lambda_2/N_f}&\equiv& \frac{1}{4\pi}N_f^2\frac{\lambda_2}{N_f}\frac{T}{m_{\rm eff}},
\label{ring2}
\end{eqnarray}
where $m_{\rm eff}=m_{\Phi}^2+bT^2/12$. It is sufficient to consider these conditions only in the symmetric phase for our analysis, since $\phi=0$ at the tricritical line. Therefore, in order to guarantee the validity of the ring-improved perturbation theory at the tricritical line, it is required that $\beta_{\lambda_1+\lambda_2/N_f},\beta_{\lambda_2/N_f}\ll 1$, while  
the perturbative expansion at zero temperature is valid  for $N_f^2(\lambda_1+\lambda_2/N_f)/(4\pi)^2\ll 1$ and $N_f\lambda_2/(4\pi)^2 \ll 1$.

%%%%%%%%%%%%%%%%%%%%%%%%%%%%%%%%%%%%%%%%%%%%%%%
\section{Analysis of the tricritical line}
\label{sec:cep}

At the critical temperature of the first-order phase transition, the effective potential satisfies the following conditions,
\begin{eqnarray}
\frac{\partial V}{\partial \phi}(\phi_c,T_c)=0,\hspace{.5cm}  
V(\phi_c,T_c)=V(0,T_c).  \label{cond-1st} 
\end{eqnarray}
A tricritical line is where $\phi_c/T_c$ is equal to zero for a finite value of $T_c$. To locate a tricritical line, we take the following approach. In the parameter region near a tricritical line, $\phi_c/T_c$ takes very small but nonzero value. In this region, we may solve Eq.(\ref{cond-1st}) by expanding the effective potential up to the sixth order of the $\phi_c/T_c$. Then we identify a tricritical line, by taking the limit as $\phi_c/T_c$ goes to zero. 

We follow this procedure by means of the ring improved one-loop finite temperature effective potential which is expanded to ${\cal O}({\cal M}^3/T^3)$ in high temperature limit. This potential is written as follows: 
\begin{align}
V_3(\phi,T)
&\equiv&T^4\left[\frac{1}{2}\left(\frac{m_{\rm eff}^2}{T^2}\right)\frac{\phi^2}{T^2}+\frac{1}{8}\left(\lambda_1+\frac{\lambda_2}{N_f}\right)\frac{\phi^4}{T^4}-\frac{1}{12\pi}\sum_{i=s}n_i \frac{{\cal M}_{i}^3(\phi,T)}{T^3}\right]. \nonumber \\
\label{effpot_pt3}
\end{align}
The high temperature expansion is valid if ${\cal M}_i/T\lesssim 1$. We evaluate later this condition at the tricritical line.
Besides the high temperature expansion, we expand ${\cal M}^3_i/T^3$ terms up to ${\cal O}(\phi^6/T^6)$,
\begin{eqnarray}
\frac{{\cal M}_i^3(\phi,T)}{T^3} &=&\frac{m_{\rm eff}^3}{T^3}\Biggl[1+\frac{3}{2}\frac{a_iT^2}{m_{\rm eff}^2}\frac{\phi^2}{T^2}
\nonumber\\
&&+\frac{3}{8}\left(\frac{a_iT^2}{m_{\rm eff}^2}\right)^2\frac{\phi^4}{T^4}-\frac{1}{16}\left(\frac{a_iT^2}{m_{\rm eff}^2}\right)^3\frac{\phi^6}{T^6}+{\cal O}\left(\frac{\phi^8}{T^8}\right)\Biggr].
\label{exp_mt3_b}
\end{eqnarray}
This expansion is valid if $\phi/T\sim0$, $m_{\rm eff}^2=m_{\Phi}^2+bT^2/12>0$ and $a_iT^2/m_{\rm eff}^2\lesssim1$. We also evaluate later this condition at the tricritical line. 
Using this expansion, Eq.(\ref{effpot_pt3}) is written as follows:
\begin{eqnarray}
V_3(\phi,T)/T^4
&=&c_0+c_2\frac{\phi^2}{T^2}+c_4\frac{\phi^4}{T^4}+c_6\frac{\phi^6}{T^6}+{\cal O}\left(\frac{\phi^8}{T^8}\right)  \label{calv2} \\
c_2&=&\frac{1}{2}\left(\frac{m_{\rm eff}^2}{T^2}-\frac{1}{4\pi}\sum_{i=s}n_ia_i\frac{m_{\rm eff}}{T}\right), \nonumber \\
c_4&=&\frac{1}{8}\left(\lambda_1+\frac{\lambda_2}{N_f}-\frac{1}{4\pi}\sum_{i=s}n_ia_i^2\frac{T}{m_{\rm eff}}\right), \nonumber \\
c_6&=&\frac{1}{192\pi}\sum_{i=s}n_ia_i^3
\frac{T^3}{m_{\rm eff}^3}. \nonumber
\end{eqnarray}
Then Eq.(\ref{cond-1st}) is solved by 
\begin{eqnarray}
\frac{\phi_c^2}{T_c^2}=\sqrt{\frac{c_2}{c_6}}=\frac{-c_4}{2c_6}  
\end{eqnarray}
for $c_2 \ge 0$, $c_4 \le 0$ and $c_6 > 0$. The tricritical line is identified in the limit as $\phi_c/T_c \searrow 0$, which means that $c_2\searrow 0,\ c_4 \nearrow 0$.\footnote{%It is necessary that $c_2$ goes to zero from above and $c_4$ goes to zero from below. It is easy to check this for our analysis. 
This result is not changed if we include the term $c_8\phi^8/T^8$, as long as $c_6, c_8>0$. }\  There are two temperatures which satisfy $c_2=0$, namely,
\begin{eqnarray}
T_1=\frac{-12m_\Phi^2}{b},\hspace{1cm}T_2=\frac{-m_\Phi^2}{(b/12-b_s^2/16\pi^2)},
\end{eqnarray}
where $b_s\equiv\sum_{i=s} n_ia_i$. From Eq.(\ref{ring1}) and Eq.(\ref{ring2}), we can see the ring-improved perturbation is broken down for $T=T_1$ at $\phi=0$, since ${\cal M}_i(0,T_1)=0$. Therefore, $T_1$ is just an artifact for our analysis, and we identify $T_2$ as the critical temperature. 
This gives the following relation for the coupling constants.
\begin{eqnarray}
\sum_{i=s}n_ia_i^2-\left(\lambda_1+\frac{\lambda_2}{N_f}\right)\sum_{i=s}n_ia_i=0.
%\sum_in_ia_i^2-\frac{m_{h,cep}^2(m_{\xi},N_f)}{v^2}\sum_in_ia_i=0.
\label{relation}
\end{eqnarray}
By means of the following relations between the coupling constants and the tree-level mass of the each fields  (see Table~\ref{tab:lsmm}):
\begin{align}
\sum_{i=s}n_ia_i&=(N_f^2+1)\frac{m_h^2(\phi_0)}{\phi_0^2}+(N_f^2-1)\frac{m_{\xi}^2(\phi_0)}{\phi_0^2},
\label{b1}
\\
\sum_{i=s}n_ia_i^2&=\left(\frac{N_f^2}{2}+2\right)\frac{m_h^4(\phi_0)}{\phi_0^4}+(N_f^2-1)\frac{m_h^2(\phi_0)m_{\xi}^2(\phi_0)}{\phi_0^4}+(N_f^2-1)\frac{m_{\xi}^4(\phi_0)}{\phi_0^4},
\label{b2}
\end{align}
we can solve Eq.(\ref{relation}) for the tree-level mass of the field $h$ as follows,
\begin{eqnarray}
m_{h,cep}(\phi_0)=2^{\frac{1}{4}}\left(\frac{N_f^2-1}{N_f^2-2}\right)^{\frac{1}{4}}m_{\xi}(\phi_0).
\label{mh0}
\end{eqnarray}
The chiral phase transition of the sigma model is first-order when $m_h< m_{h,cep}$. The relation Eq.(\ref{relation}) means that $m_{h,cep}$ becomes smaller for smaller $m_\xi$ and approaches a finite value for large $N_f$. Note that $m_\Phi^2$ is canceled out in the above calculation. This means that this result is independent of the stationary condition of the zero temperature symmetry breaking.\footnote{The truncation of the logarithmic terms does not mean we analyze the tricritical line with the tree-level potential. This terms is just {\it suppressed} as ${\cal O}({\cal M}^4/T^4)$ and we can  consistently neglect them in our analysis up to ${\cal O}({\cal M}^3/T^3)$. At lower temperature, the effective potential should be modified, for example, to solve the stationary condition at zero temperature. Our claim that the tricritical line is not affected by $m_\Phi$ is meaningful in this point.}\ Also note that the contribution from the fermions to the tricritical line of ${\cal O}(y^2)$ is canceled out and the possible leading contribution is ${\cal O}(y^4)$. 

\subsubsection*{The validity of the perturbation theory}

Although we assumed the perturbation theory for the above argument, it is necessary to examine its validity. We give here the argument about the validity of the zero temperature perturbation theory, the high temperature expansion, the order parameter expansion and the ring-improved perturbation theory at the tricritical line. 

The conditions for the validity of these approximations are summarized as follows:

\

\noindent zero temperature perturbation theory:
\begin{eqnarray}
\frac{b_s}{4\pi}\lesssim1,\hspace{1cm}
\frac{1}{4\pi}n_\psi y^2\lesssim1.
\label{zero}
\end{eqnarray}
high temperature expansion:
\begin{eqnarray}
\frac{\mathcal{M}_{i=s,f}}{T}\lesssim1.
\label{highT}
\end{eqnarray}
order parameter expansion:
\begin{eqnarray}
\frac{a_i}{m_\Phi^2/T^2+b/12}\lesssim1.
\label{order}
\end{eqnarray}
ring-improved perturbation theory:
\begin{eqnarray}
\beta_{\lambda_1+\lambda_2/N_f}&=& \frac{1}{4\pi}\left(\lambda_1+\frac{\lambda_2}{N_f}\right)\frac{T}{m_{\rm eff}}\ll1 ,
\label{ring1-2}\\
\beta_{\lambda_2/N_f}&=& \frac{1}{4\pi}N_f^2\frac{\lambda_2}{N_f}\frac{T}{m_{\rm eff}}\ll1.
\label{ring2-2}
\end{eqnarray}
To justify our calculation of the tricritical line, all of these conditions must be satisfied in the limit as $\phi_c/T_c$ goes to zero. 
%In  the following evaluation, we assume the validity of the

At first, we examine the condition for the high temperature expansion, Eq.(\ref{highT}). Since $T=T_2$ at the tricritical line, for scalar bosons, it follows 
\begin{eqnarray}
\frac{\mathcal{M}_{s,i}}{T_c}\to \frac{b_s}{4\pi} \ \ \ \ (\phi_c/T_c\to0) .
\label{c2-0}
\end{eqnarray}
Therefore, if the zero temperature perturbation theory is valid: $b_s/4\pi\lesssim 1$, it follows $\mathcal{M}_i/T_c\lesssim 1$ and high temperature expansion is also valid. The high temperature expansion for the fermions are valid in the limit as $\phi_c/T_c$ goes to zero, since ${\cal M}_\psi/T_c$ is proportional to $\phi_c/T_c$.

Next, we examine the condition for the order parameter expansion, Eq.(\ref{order}). At the tricritical line, this factor is evaluated as
\begin{eqnarray}
\frac{a_i}{m_\Phi^2/T^2+b/12}\to a_i\frac{16\pi^2}{b_s^2}\ \ \ \ (\phi_c/T_c\to0)
\end{eqnarray}
Therefore, in order to satisfy the condition Eq.(\ref{order}), it must be $N_f\gg1$ and $a_i\ll1$ and $b_s/4\pi\lesssim 1$. 

At last, we examine the conditions for the ring-improved perturbation. In the limit as $\phi_c/T_c$ goes to zero,
\begin{eqnarray}
\frac{T_c}{\mathcal{M}_i}&\to&\frac{4\pi}{b_s} \nonumber \\
&\sim&\frac{1}{N_f^2}\frac{4\pi}{(\lambda_1+\lambda_2/N_f)+\lambda_2/N_f} \ \ \ \ \ \ (N_f\gg1) .
\end{eqnarray}
Then
\begin{eqnarray}
\beta_{\lambda_1+\lambda_2/N_f}
&\sim&\frac{1}{N_f^2}\frac{\lambda_1+\lambda_2/N_f}{(\lambda_1+\lambda_2/N_f)+\lambda_2/N_f} \ \ \ \ \ \ (N_f\gg1) ,
\label{ring_validity_1} \\
\beta_{\lambda_2/N_f}
&\sim&\frac{\lambda_2/N_f}{(\lambda_1+\lambda_2/N_f)+\lambda_2/N_f} \ \ \ \ \ \ (N_f\gg1) .
\label{ring_validity_2}
\end{eqnarray}
Therefore it follows that the ring-improved perturbation theory is valid if $N_f\gg1$ and $(\lambda_1+\lambda_2/N_f)\gg \lambda_2/N_f$. Concerning this evaluation, we note that at the tricritical line, the rhs of Eq.(\ref{ring_validity_2}) is evaluated as $\beta_{\lambda_2/N_f}\sim 0.4$. Since this value is not $\ll {\cal O}(1)$, but $\lesssim {\cal O}(1)$, we need a more precise argument to confirm our result.\footnote{In our evaluation, $\beta_{\lambda_2/N_f}$ is divided by $4\pi$ compared to the ordinary criteria of the ring improvement as, \cite{Quiros:1999jp} \ because we should take into account the loop factor in our argument. In \cite{Quiros:1999jp} loop factor is neglected, for example, as $m_{eff}^2\sim\lambda^2T^2$. This corresponds to our case with neglecting factor 1/4$\pi$ in Eq.(\ref{c2-0}). We also consider the numerical factor which is neglected in the ordinary argument, so our argument is not so crude.}   

These result is summarized as follows. Our analysis is valid if $a_i\ll1$ and $N_f\gg1$ and   $(\lambda_1+\lambda_2/N_f)\gg \lambda_2/N_f$. This parameter region seems to be small, but there is a region where our evaluations is reliable. The parameter region that $N_f\gg1$ and $a_i\ll1$ and $b_s/4\pi\lesssim 1$ is just the parameter region where we can analyze with the large $N_f$ expansion.

%%%%%%%%%%%%%%%%%%%%%%%%%%%%%%%%%%%%%%%%%%%%%%%
\section{Summary and Discussion}
\label{sec:sd}

% summary
We have studied the tricritical line of the U($N_f$)$\times$U($N_f$) sigma model with Yukawa interaction in an analytic approach by means of the ring improved one-loop finite temperature effective potential. To evaluate the tricritical line analytically, we have expanded the effective potential up to ${\cal O}(M^3/T^3)$ in the high temperature expansion, and up to ${\cal O}(\phi^6/T^6)$ in the order parameter expansion. In this approximation, we have shown that the tricritical line can be examined to the lowest order of the coupling constants and it follows a analytic relation between the tree-level masses of the scalar bosons. This analysis is valid in the parameter region where $a_i\ll1$ and $N_f\gg1$ and  $(\lambda_1+\lambda_2/N_f)\gg \lambda_2/N_f$. 

% with gauge bosons
We can examine the contribution of gauge bosons by means of our technique. However, in this case, since the parameter to evaluate the validity of the ring improvement for the transverse modes of the gauge bosons, $\beta_{gT}=g^2T/M^2_{gT}$, diverges for the small value of $\phi/T$, we cannot perform a reliable analysis near the tricritical line.\footnote{We find there is no tricritical line in this case because of the $\phi^3$ term from the transverse modes of the gauge bosons.}\ As a prescription, we could add the magnetic mass term $\sim g^2T$ to the transverse modes which reduce the singularity of the $\beta_{gT}$. However, in this case, we find that the coefficient of the order parameter expansion is proportional to $1/g^2$, therefore the perturbation is not valid after all. We need another prescription to take care of this problem.

% next-to-leading order of the coupling constants
Although we have expanded the potential up to ${\cal O}(M^3/T^3)$ in the high temperature limit, we can improve this approximation by including the terms of ${\cal O}(M^4/T^4)$. Since these terms contain the $\log(T^2/\mu^2)$ terms, we cannot solve analytically the conditions of the first-order phase transition for the critical temperature. Therefore we need another prescription to derive the analytic relation between the coupling constants on the tricritical line up to the next-to-leading-order. 
%If we can solve this problem, the tree-level mass relation will be improve to the next-leading-order of the coupling constants. 

% applications to the phenomenology
% a topic of the O(N) sigma model
By means of our technique, we might study the many of phenomenological issues, such as QCD chiral phase transition, electroweak phase transition or the other problems of the cosmology. However, in order to address these issues, we need to consider a lot of problems, such as the effect of the finite density or the infrared singularity of the gauge bosons as mentioned above etc. 
%However our technique is too poor in the two point to discuss these issues. The one is the problem of the infrared singularities of transverse gauge bosons as mentioned before and the two is the narrow parameter region where our technique is valid. 
%To apply our technique to the electroweak phase transition of the standard model, we need to analyze the O(4) sigma model as a Higgs sector. Although this model has been already examined as the O($N$) sigma model in the large $N$ limit by the many of authors \cite{Arnold:1993bq}, our technique could add some insight to their results.
It is also interesting to compare our analysis to the results of the numerical simulations, and to examine the non-perturbative feature of the tricritical line.

%%%%%%%%%%%%%%%%%%%%%%%%%%%%%%%%%%%%%%%%%%%%%%%

\subsection*{Acknowledgments}
The author would like to thank Y.~Kikukawa, M.~Kohda 
for valuable discussions.

%#################################################################

\end{document}